\documentclass[a4paper,twoside]{article}
\newcommand{\affil}[1]{$^{\rm #1}$}
\usepackage[authoryear]{natbib}
\bibpunct{(}{)}{;}{a}{}{,}
\usepackage{graphicx}
\usepackage{epstopdf}
\date{} 
\renewcommand{\arraystretch}{1.2}

\usepackage{url}
\usepackage{subfigure}
\usepackage{deluxetable}
\usepackage{pdflscape}

\title{\large\bf\flushleft Adaptive Optics Simulations for Siding Spring}
\author{\parbox{\textwidth}{\flushleft
\vspace{-0.5cm}
{\it Michael Goodwin\affil{A,C,D}, Charles Jenkins\affil{A,E} and Andrew Lambert\affil{B} } \\ 
\vspace{0.4cm}
{\small \affil{A}\,Research School of Astronomy  Astrophysics, Australian National University, Mt Stromlo Observatory, via Cotter Rd, Weston, ACT 2611, Australia}\\
{\small \affil{B}\,School of Engineering and Information Technology, UNSW@ADFA}\\
{\small \affil{C}\,Corresponding author. Email: mgoodwin@aao.gov.au}\\
{\small \affil{D}\,Current address: Australian Astronomical Observatory, PO Box 296, Epping, NSW 1710, Australia }\\
{\small \affil{E}\,Current address: Earth Science and Resource Engineering, CSIRO}}}

\begin{document}
\twocolumn[
\begin{changemargin}{.8cm}{.5cm}
\begin{minipage}{.9\textwidth}
\vspace{-1cm}
\maketitle
%
%
\small{\bf Abstract:}

Using an observational derived  model optical turbulence profile (model-OTP) we have investigated the performance of Adaptive Optics (AO) at Siding Spring Observatory (SSO), Australia. The simulations cover the performance for AO techniques of single conjugate adaptive optics (SCAO), multi-conjugate adaptive optics (MCAO) and ground-layer adaptive optics (GLAO). The simulation results presented in this paper predict the performance of these AO techniques as applied to the Australian National University (ANU) 2.3~m and Anglo-Australian Telescope (AAT) 3.9~m telescopes for astronomical wavelength bands J, H and K. The results indicate that AO performance is best for the longer wavelengths (K-band) and in the best seeing conditions (sub 1-arcsecond). The most promising results are found for GLAO simulations (field of view of 180 arcsecs), with the field RMS for encircled energy 50$\%$ diameter (EE50d) being uniform and minimally affected by the free-atmosphere turbulence. The GLAO performance is reasonably good over the wavelength bands of J, H and K. The GLAO field mean of EE50d is between 200~mas to 800~mas, which is a noticeable improvement compared to the nominal astronomical seeing (870 to 1700~mas).


\medskip
\medskip
\end{minipage}
\end{changemargin}
]
\small

\section{Introduction}
Of interest is the performance of Adaptive Optics (AO) at Siding Spring Observatory (SSO). It could be that the installation of AO for the 2.3~m ANU and  3.9~m AAT may open the door for new science programs and discoveries. Certain AO correction modes may provide encouraging performance gains at SSO, despite the relatively moderate seeing conditions. It is therefore important to ascertain the  performance predictions for AO. In addition, AO systems are becoming more achievable and affordable for most astronomical observatories (not the case when adaptive optics was first envisioned by Horace W. Babcock in 1953~\citep{Hardy1998}). Evidence of this fact is the success of adaptive optics demonstrated with the modern 8-10~m class telescopes. Good performance has been reported with the 10~m Keck II Telescope~\citep{vanDam2006}, the 8~m Very Large Telescope~\citep{Rousset2003}, the 8.2~m Subaru Telescope~\citep{Iye2004} and 8~m Gemini North Telescope~\citep{Stoesz2004} and others.

To predict the performance of AO at SSO requires the characterisation and modeling of the atmospheric optical turbulence profile (model-OTP) based on observational results. The observational results for turbulence profiling at SSO and model-OTP are reported in the paper 'Characterisation of the Optical Turbulence at Siding Spring' ~\citep{Goodwin2012} and in PhD thesis by ~\cite{Goodwin2009}. Adaptive optics simulations use the site-characteristic model of the optical turbulence profile, or model-OTP, to predict the performance of various adaptive optics technologies.

The purpose of adaptive optics for astronomical telescopes is to remove the wavefront aberrations from the optical path between the science object and the imaging detector. When this is successful the quality of the image is limited by the diffraction limit of the astronomical telescope. Most of the wavefront aberrations are induced by the atmospheric turbulence as random phase perturbations within the beam path. An adaptive optics system attempts to measure these phase perturbations and correct (e.g. spatial phase modulator) them in real-time, typically on timescales of milliseconds, to restore image quality.

Adaptive optics improves the performance of most optical instruments, including spectrographs, interferometers and imaging detectors. Adaptive optics can also compensate for telescope tracking errors and wind buffeting, as well as slow timescale aberrations such as mirror/dome seeing and mirror gravity distortions. Low frequency errors are the largest errors and are controlled by a separate system known as ``active optics".

A good introduction the subject of AO can be found in the book 'Adaptive Optics for Astronomical Telescopes' by \cite{Hardy1998}. An overview of AO and the various correction modes can be found in the author's PhD thesis titled 'Turbulence profiling at Siding Spring and Las Campanas Observatories' ~\citep{Goodwin2009}. 

This paper discusses the YAO numerical simulation code (authored by Francis Rigaut~\citep{Rigaut2007}) and corresponding performance predictions for the 2.3~m ANU and  3.9~m AAT at SSO. Section 2, 3 and 4 discusses the turbulence model, simulation tool and simulation configurations. Section 5 reports the simulation results. Concluding remarks are provided in section 6.

\section{Turbulence Model}

\label{sec:turbulencemodel}

The simulation code uses the SSO model-OTP  in the paper 'Characterisation of the Optical Turbulence at Siding Spring' ~\cite{Goodwin2012}. The model-OTP is a statistical thin-layer characterization, based on measurements of the turbulence profile above SSO conducted over years 2005 and 2006. The simulation code use  fractional layer strengths  and are listed in Table~\ref{tab:modelotp_ssoall_summary_fractional}. The simulation code  uses the corresponding  layer wind speeds and directions by~\cite{Goodwin2012}. From Table~\ref{tab:modelotp_ssoall_summary_fractional}, it is evident that the  ground-layer contains the bulk fraction of the turbulence integral. It is noted by~\cite{Goodwin2012} and ~\cite{Goodwin2009} that the free-atmosphere ($>$ 500~m) is comparable to the 'good' seeing astronomical sites, like Cerro Pachon, Chile.

\begin{landscape}
\begin{table}
\center
\begin{tabular}{|l|l|lll|lll|lll|}
\cline{3-11}
\multicolumn{1}{l}{} &  & \multicolumn{9}{c|}{Model Turbulence Profiles ($J$, Fractional) - SSO (Run 1-8: May 2005 to June 2006)} \\
\cline{2-11}
\multicolumn{1}{l|}{} & GL & \multicolumn{3}{c|}{Good} & \multicolumn{3}{c|}{Typical} & \multicolumn{3}{c|}{Bad} \\
\cline{2-11}
\multicolumn{1}{l|}{} & FA & \multicolumn{1}{l|}{Good} & \multicolumn{1}{l|}{Typical} & Bad & \multicolumn{1}{l|}{Good} & \multicolumn{1}{l|}{Typical} & Bad & \multicolumn{1}{l|}{Good} & \multicolumn{1}{l|}{Typical} & Bad \\
\hline
Parameter & Units & \multicolumn{1}{l|}{1} & \multicolumn{1}{l|}{2} & 3 & \multicolumn{1}{l|}{4} & \multicolumn{1}{l|}{5} & 6 & \multicolumn{1}{l|}{7} & \multicolumn{1}{l|}{8} & 9 \\
\hline
37.5 & / 1.0 & 0.8810 & 0.7635 & 0.6139 & 0.8272 & 0.7575 & 0.6564 & 0.6901 & 0.6509 & 0.5897 \\
250 & / 1.0 & 0.0464 & 0.0402 & 0.0324 & 0.1294 & 0.1185 & 0.1027 & 0.2815 & 0.2655 & 0.2405 \\
1000 & / 1.0 & 0.0451 & 0.0773 & 0.0870 & 0.0270 & 0.0489 & 0.0592 & 0.0176 & 0.0329 & 0.0417 \\
3000 & / 1.0 & 0.0059 & 0.0708 & 0.1483 & 0.0035 & 0.0447 & 0.1010 & 0.0023 & 0.0301 & 0.0711 \\
6000 & / 1.0 & 0.0038 & 0.0150 & 0.0467 & 0.0023 & 0.0095 & 0.0318 & 0.0015 & 0.0064 & 0.0224 \\
9000 & / 1.0 & 0.0054 & 0.0140 & 0.0289 & 0.0033 & 0.0088 & 0.0197 & 0.0021 & 0.0060 & 0.0139 \\
13500 & / 1.0 & 0.0123 & 0.0192 & 0.0428 & 0.0073 & 0.0122 & 0.0292 & 0.0048 & 0.0082 & 0.0206 \\
\hline
37.5 & m/s & 2.1981 & 2.1981 & 2.1981 & 5.6605 & 5.6605 & 5.6605 & 9.4200 & 9.4200 & 9.4200 \\
250 & m/s & 2.2506 & 2.2506 & 2.2506 & 5.7810 & 5.7810 & 5.7810 & 9.6142 & 9.6142 & 9.6142 \\
1000 & m/s & 2.5495 & 6.3707 & 10.4871 & 2.5495 & 6.3707 & 10.4871 & 2.5495 & 6.3707 & 10.4871 \\
3000 & m/s & 5.1620 & 9.9300 & 14.7605 & 5.1620 & 9.9300 & 14.7605 & 5.1620 & 9.9300 & 14.7605 \\
6000 & m/s & 19.0935 & 23.4552 & 27.9741 & 19.0935 & 23.4552 & 27.9741 & 19.0935 & 23.4552 & 27.9741 \\
9000 & m/s & 32.0000 & 38.6276 & 43.7936 & 32.0000 & 38.6276 & 43.7936 & 32.0000 & 38.6276 & 43.7936 \\
13500 & m/s & 10.4619 & 26.4419 & 41.6250 & 10.4619 & 26.4419 & 41.6250 & 10.4619 & 26.4419 & 41.6250 \\
\hline
$J_{GL}$ & $10^{-13} m^{1/3}$ & 4.9614 & 4.9614 & 4.9614 & 8.5562 & 8.5562 & 8.5562 & 13.2849 & 13.2849 & 13.2849 \\
$J_{FA}$ & $10^{-13} m^{1/3}$ & 0.3880 & 1.2115 & 2.7159 & 0.3880 & 1.2115 & 2.7159 & 0.3880 & 1.2115 & 2.7159 \\
$J$ & $10^{-13} m^{1/3}$ & 5.3493 & 6.1729 & 7.6773 & 8.9442 & 9.7678 & 11.2721 & 13.6728 & 14.4964 & 16.0008 \\
\hline
$F_{GL}$ & / 1.0 & 0.9275 & 0.8037 & 0.6462 & 0.9566 & 0.8760 & 0.7591 & 0.9716 & 0.9164 & 0.8303 \\
$F_{FA}$ & / 1.0 & 0.0725 & 0.1963 & 0.3538 & 0.0434 & 0.1240 & 0.2409 & 0.0284 & 0.0836 & 0.1697 \\
\hline
$\epsilon_{GL}$ & arcsecs & 0.8277 & 0.8277 & 0.8277 & 1.1478 & 1.1478 & 1.1478 & 1.4945 & 1.4945 & 1.4945 \\
$\epsilon_{FA}$  & arcsecs & 0.1794 & 0.3552 & 0.5766 & 0.1794 & 0.3552 & 0.5766 & 0.1794 & 0.3552 & 0.5766 \\
$\epsilon$ & arcsecs & 0.8659 & 0.9436 & 1.0755 & 1.1787 & 1.2427 & 1.3542 & 1.5206 & 1.5749 & 1.6710 \\
\hline
$\theta_0$ & arcsecs & 6.4233 & 3.7172 & 2.0123 & 6.3684 & 3.7043 & 2.0098 & 6.2255 & 3.6700 & 2.0030 \\
$\tau$ & ms & 11.7922 & 5.3516 & 2.3291 & 4.5855 & 3.5038 & 2.0310 & 2.2067 & 2.0112 & 1.5242 \\
\hline
Probability & / 1.0 & 0.0625 & 0.1250 & 0.0625 & 0.1250 & 0.2500 & 0.1250 & 0.0625 & 0.1250 & 0.0625 \\
\hline
\end{tabular}
\caption{\label{tab:modelotp_ssoall_summary_fractional} Tabulated values for the final model-OTP for the SSO (Run 1-8: May 2005 to June 2006), with layers specified as fractional amount of total turbulence integral, $J$, with corresponding wind speed, $m/s$  ~\citep{Goodwin2012,Goodwin2009}.}
\end{table}
\end{landscape}

\section{Simulation Code}

The simulation code used to perform the adaptive simulations is an open source numerical simulation code called YAO, written by Francis Rigaut~\citep{Rigaut2007}. The YAO simulation code is a Monte-Carlo adaptive optics simulation tool coded in YORICK~\citep{Munro2005}. YORICK is an open source interpreted programming language for scientific simulation codes. The YAO simulation code has custom developed functions to simulate the wavefront sensor (WFS), the deformable mirror (DM) and other aspects of an adaptive optics loop~\citep{Rigaut2007}. The YAO simulation code is provided with a set of example scripts which can be easily modified to simulate SCAO, MCAO and GLAO for a specific model atmosphere, telescope and adaptive optics system configuration. The YAO simulation code was selected based on its extensive functionality, ease of use (documentation and examples), as well as being open source software. The YAO simulation code has been verified with independent code for the case of SCAO~\citep{Goodwin2009}.

\section{Simulation Configurations}

The simulation results are based on the adaptive optics correction modes of SCAO, MCAO and GLAO for the ANU 2.3~m and AAT 3.9~m using the YAO adaptive optics numerical simulation code. Each adaptive optics correction mode and telescope specification is associated with a set of simulation input configuration parameters and system geometry.

\begin{figure*}[htbp]
  \begin{center}
    \mbox{
      \subfigure[]{\scalebox{1.0}{\includegraphics[width=0.45\textwidth]{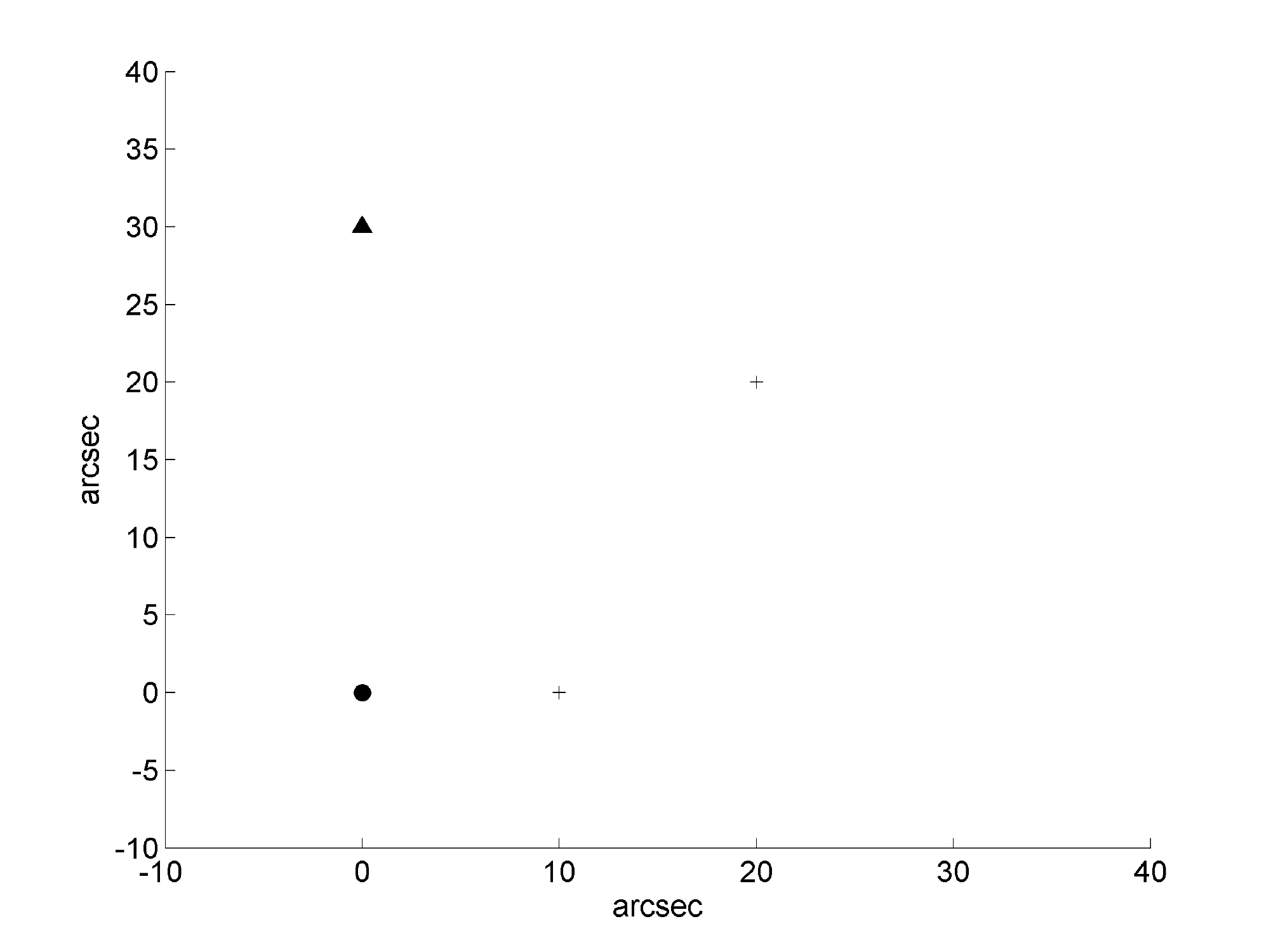}}} \quad
      \subfigure[]{\scalebox{1.0}{\includegraphics[width=0.45\textwidth]{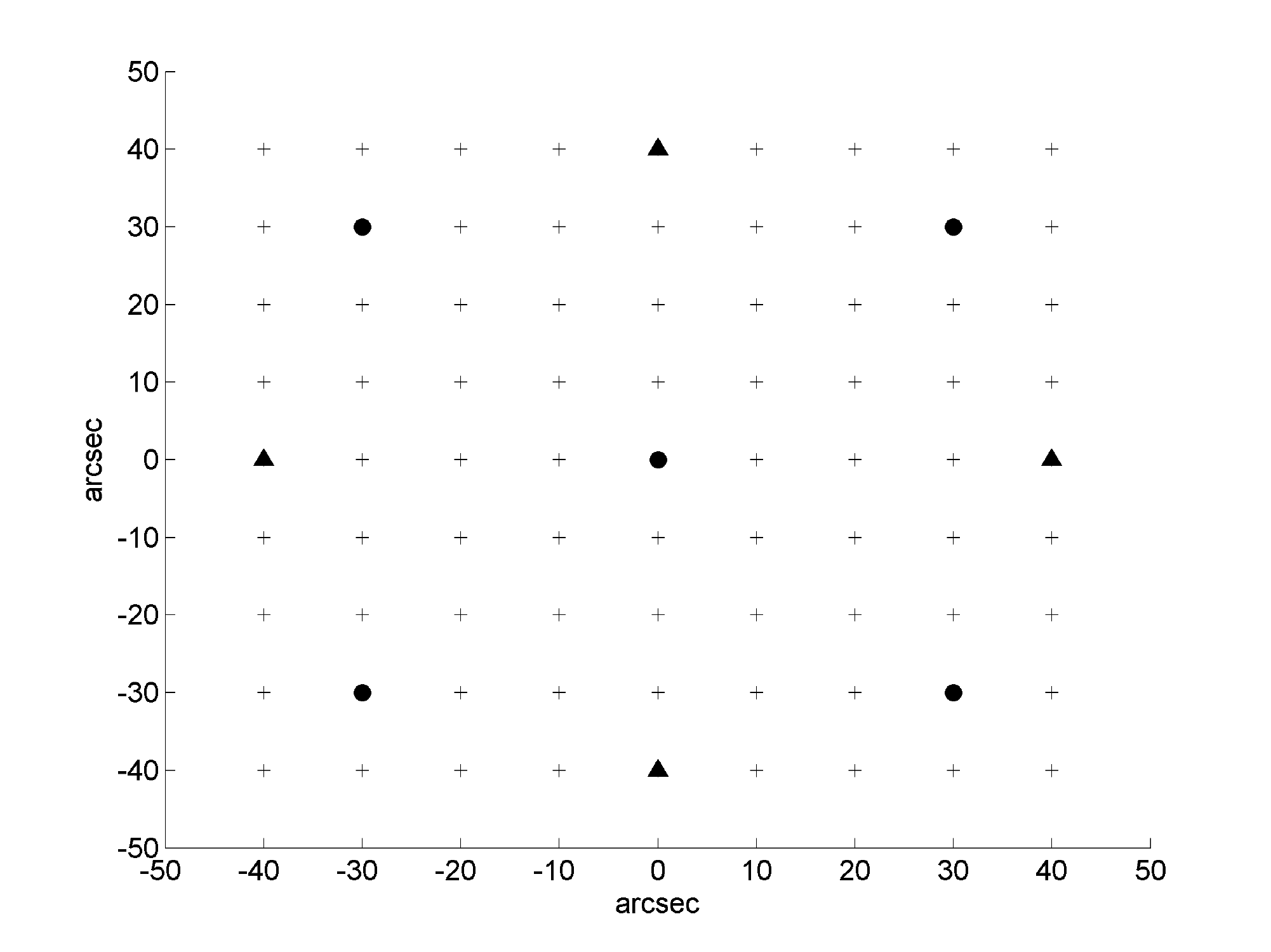}}}
      }
    \mbox{
      \subfigure[]{\scalebox{1.0}{\includegraphics[width=0.45\textwidth]{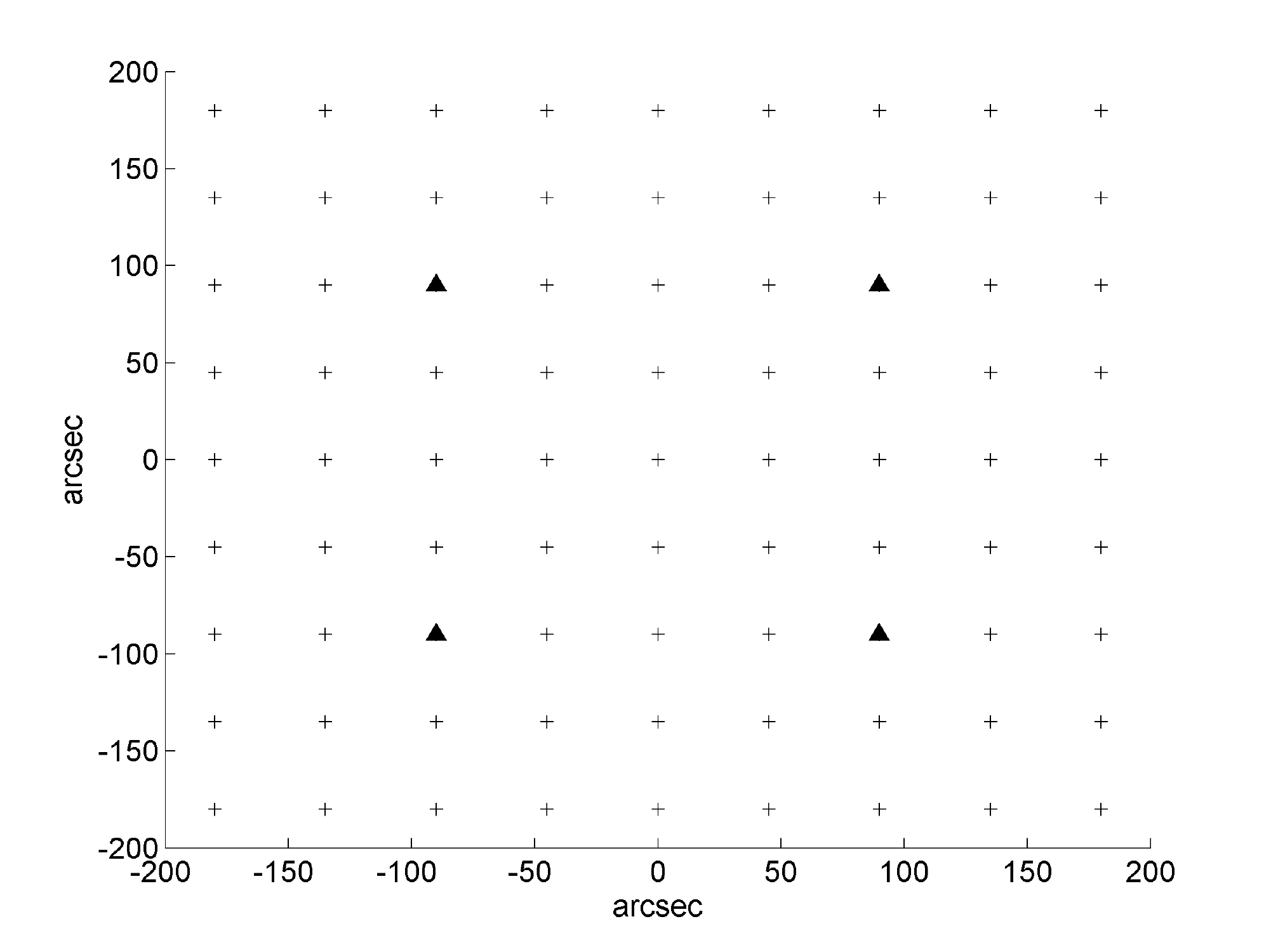}}} \quad
      }
    \caption[Geometry of configurations for SCAO, MCAO and GLAO as used in adaptive optic simulations.] {Geometry of configurations for (a) SCAO (b) MCAO and (c) GLAO configurations as used in adaptive optics simulations for the ANU 2.3~m and AAT 3.9~m  telescopes. The NGS are marked with filled triangles; LGS are marked with filled circles and Field stars (FS) as plus signs. } \label{fig:sim_config_geometry}
  \end{center}
\end{figure*}

The schematic for the system geometry of the SCAO, MCAO and GLAO adaptive optics models as used with the ANU 2.3~m and AAT 3.9~m telescopes are shown in Figure~\ref{fig:sim_config_geometry}. The large number of field stars (FS) in Figure~\ref{fig:sim_config_geometry}~(b) and (c) are artificial and inserted onto a regular grid to calculate the Strehl, FWHM and EE50d performance parameters as a function of field location.

The input configuration parameters for the SCAO adaptive optics model as used with the ANU 2.3~m and AAT 3.9~m  telescopes are tabulated in Table~\ref{tab:aosim_scao_parameters}. The sky coverage for the SCAO adaptive optics model, given a NGS (tip-tilt) magnitude within a search radius of 30" from LGS (science target), as used with the ANU 2.3~m and AAT 3.9~m telescopes, is tabulated in Table~\ref{tab:aosim_scao_skycoverage}, and were calculated by~\cite{Chun2000}.

The input configuration parameters for the MCAO adaptive optics model as used with the ANU 2.3~m and AAT 3.9~m  are tabulated in Table~\ref{tab:aosim_mcao_parameters}.

The input configuration parameters for the GLAO adaptive optics model as used with the ANU 2.3~m and AAT 3.9~m are tabulated in Table~\ref{tab:aosim_glao_parameters}.

The input parameters, e.g. mirror type and transmission, were in part adapted from the test examples that were distributed with YAO given the complexity involved in the simulations. The hardware simulation choices are based on readily available technology at the time of simulation (around the year 2008/2009). The parameters were sensible and kept consistent amongst the correction modes, e.g. for SCAO, a  NGS of 15 mags for tip-tilt sensing seems appropriate as performance starts to fall-off with fainter tip-tilt stars.

\begin{table}[h]
\begin{center}
\caption{Estimated sky coverage for SCAO based on NGS (tip-tilt) search radius of 30" from LGS (science target).}\label{tab:aosim_scao_skycoverage}
\begin{tabular}{llll}
\hline Tilt NGS $m_R$ & 15 & 17 & 19  \\
\hline Galactic Latitude 30 & 12\% & 35\% & 73\% \\
Galactic Latitude 90 & 4\% & 12\% & 28\% \\
\hline
\end{tabular}
\medskip\\
\end{center}
\end{table}

\newpage
\onecolumn

\begin{deluxetable}{lll}
\tabletypesize{\footnotesize}
\tablecolumns{8}
\tablewidth{0pt}
\tablecaption{\label{tab:aosim_scao_parameters} Parameters of the models used in SCAO simulation codes.}
\tablehead {
\colhead {Parameter (SCAO)} &
\colhead {ANU 2.3m} &
\colhead {AAO 3.9m} }
\startdata

Model-OTP & SSO &  SSO \\
\hline
LGS (high order source) &  &    \\
LGS Position (x,y) & (0",0") & (0",0")  \\
LGS Power & 10W & 10W \\
LGS Sodium Layer Altitude & 95,000m & 95,000m  \\
LGS Sodium Layer Thickness & 8,000m & 8,000m  \\
LGS Return (photons/cm2/s/W) & 30 & 30  \\
\hline
NGS (tip/tilt source) &  &   \\
NGS  Position (x,y) & (0",30") & (0",30")  \\
NGS mR & 15 (17,19) & 15 (17,19) \\
\hline
WFS (high order sensing) & LGS & LGS  \\
WFS (Shack Hartmann) & 11x11 & 14x14  \\
WFS Subaperture (Pixels) & 8x8 & 8x8  \\
WFS Pixel Scale & 0.5" / pixel  & 0.5" / pixel  \\
WFS Read Noise RMS (e/pixel) & 2 & 2 \\
WFS Frame Rate & 500 fps & 500 fps \\
WFS Wavelength & 589nm & 589nm  \\
\hline
WFS (tip/tilt sensing) & NGS & NGS  \\
WFS (Shack Hartmann, APD) & 1x1 & 1x1  \\
WFS Subaperture (Pixels) & 4x4 & 4x4  \\
WFS Pixel Scale & 0.5" / pixel  & 0.5" / pixel  \\
WFS Read Noise RMS (e/pixel) & 0 & 0  \\
WFS Frame Rate & 50 fps & 50 fps  \\
WFS Wavelength & 700nm & 700nm  \\
\hline
DM Conjugate Height & 0m & 0m  \\
DM Actuators (DOF) & 113 & 177 \\
\hline
Field Stars &  &    \\
FS 1 Position (x,y) & (0",0") & (0",0") \\
FS 2 Position (x,y) & (0",10") & (0",10")  \\
FS 3 Position (x,y) & (20",20") & (20",20") \\
FS Wavelengths (microns) & 1.2, 1.65, 2.2 & 1.2, 1.65, 2.2  \\
FS Integration Time & 2s & 2s \\
FS Zenith Angle & 0 & 0 \\
\enddata
\end{deluxetable}

\begin{deluxetable}{llll}
\tabletypesize{\footnotesize}
\tablecolumns{8}
\tablewidth{0pt}
\tablecaption{\label{tab:aosim_mcao_parameters} Parameters of the models used in MCAO simulation code.}
\tablehead {
\colhead {Parameter (MCAO)} &
\colhead {ANU 2.3m} &
\colhead {AAO 3.9m}  }
\startdata

Model-OTP & SSO &  SSO \\
\hline
LGS (high order source) &  &   \\
LGS 1 Position (x,y) & (-30",-30") & (-30",-30") \\
LGS 2 Position (x,y) & (30",-30") & (30",-30") \\
LGS 3 Position (x,y) & (0",0") & (0",0")  \\
LGS 4 Position (x,y) & (-30",30") & (-30",30")  \\
LGS 5 Position (x,y) & (30",30") & (30",30")  \\
LGS Power & 10W & 10W \\
LGS Sodium Layer Altitude & 95,000m & 95,000m \\
LGS Sodium Layer Thickness & 8,000m & 8,000m  \\
LGS Return (photons/cm2/s/W) & 30 & 30  \\
\hline
NGS (tip/tilt source) &  &   \\
NGS  1 Position (x,y) & (40",0") & (40",0")  \\
NGS  2 Position (x,y) & (0",-40") & (0",-40")  \\
NGS  3 Position (x,y) & (-40",0") & (-40",0") \\
NGS  4 Position (x,y) & (0",40") & (0",40")  \\
NGS mR & 5 & 5  \\
\hline
WFS (high order sensing) & LGS & LGS  \\
WFS (Shack Hartmann) & 11x11 & 14x14 \\
WFS Subaperture (Pixels) & 4x4 & 4x4  \\
WFS Pixel Scale & 0.5" / pixel  & 0.5" / pixel \\
WFS Read Noise RMS (e/pixel) & 2 & 2 \\
WFS Frame Rate & 500 fps & 500 fps \\
WFS Wavelength & 589nm & 589nm  \\
\hline
WFS (tip/tilt sensing) & NGS & NGS \\
WFS (Shack Hartmann, APD) & 1x1 & 1x1  \\
WFS Subaperture (Pixels) & 4x4 & 4x4  \\
WFS Pixel Scale & 0.5" / pixel  & 0.5" / pixel \\
WFS Read Noise RMS (e/pixel) & 0 & 0  \\
WFS Frame Rate & 50 fps & 50 fps  \\
WFS Wavelength & 700nm & 700nm  \\
\hline
DM 1 Conjugate Height & 0m & 0m \\
DM 1 Actuators (DOF) & 113 & 177  \\
DM 2 Conjugate Height & 2000m & 2000m \\
DM 2 Actuators (DOF) & 201 & 380  \\
DM 3 Conjugate Height & 7000m & 7000m  \\
DM 3 Actuators (DOF) & 79 & 201  \\
\hline
Field Stars &  &   \\
FS Position (x,y)  &  &    \\
(0",0")- (40",40") & 5x5 grid & 5x5 grid  \\
FS Wavelengths (microns) & 1.2, 1.65, 2.2 & 1.2, 1.65, 2.2 \\
FS Integration Time & 1s & 1s \\
FS Zenith Angle & 0 & 0  \\
\enddata
\end{deluxetable}

\begin{deluxetable}{llll}
\tabletypesize{\footnotesize}
\tablecolumns{8}
\tablewidth{0pt}
\tablecaption{\label{tab:aosim_glao_parameters} Parameters of the models used in GLAO simulation code.}
\tablehead {
\colhead {Parameter (GLAO)} &
\colhead {ANU 2.3m} &
\colhead {AAO 3.9m}  }
\startdata
NGS (high order, tip/tilt source) &  &   \\
NGS  1 Position (x,y) & (90",90") & (90",90") \\
NGS  2 Position (x,y) & (90",-90") & (90",-90") \\
NGS  3 Position (x,y) & (-90",-90") & (-90",-90") \\
NGS  4 Position (x,y) & (-90",90") & (-90",90") \\
NGS mR & 11 & 11  \\
\hline
WFS (high order sensing) & NGS & NGS  \\
WFS (Shack Hartmann) & 11x11 & 18x18 \\
WFS Subaperture (Pixels) & 4x4 & 4x4 \\
WFS Pixel Scale & 0.75" / pixel  & 0.5" / pixel  \\
WFS Read Noise RMS (e/pixel) & 2 & 2 \\
WFS Frame Rate & 200 fps & 200 fps  \\
WFS Wavelength & 700nm & 700nm \\
\hline
DM Conjugate Height & 0m & 0m  \\
DM Type & Bimorph & Bimorph  \\
\hline
Field Stars &  &    \\
FS Position (x,y)  &  &   \\
(-180", -180")- (180",180") & 9x9 grid & 9x9 grid \\
FS Wavelengths (microns) & 1.2, 1.65, 2.2 & 1.2, 1.65, 2.2  \\
FS Integration Time & 1.25s & 1.25s \\
FS Zenith Angle & 0 & 0  \\
\enddata
\end{deluxetable}

\twocolumn


\section{Simulation Results}

This section reports the results of adaptive optics techniques SCAO, MCAO and GLAO as applied to the ANU 2.3~m and AAT 3.9~m telescopes for wavelength bands J, H and K using the SSO model-OTP as  tabulated in Table~\ref{tab:modelotp_ssoall_summary_fractional}, see \S\ref{sec:turbulencemodel}.

\subsection{SCAO Results}

The input SCAO configuration parameters as used in the YAO simulation code for the ANU 2.3~m are tabulated in Table~\ref{tab:aosim_scao_parameters} with a schematic of the geometry shown in Figure~\ref{fig:sim_config_geometry}. Likewise for the AAT 3.9~m the parameters are tabulated in Table~\ref{tab:aosim_scao_parameters} and the geometry shown in Figure~\ref{fig:sim_config_geometry}. 

\begin{figure}[htbp]
\centering
  \includegraphics[width=1.0\columnwidth, bb=0 0  420 560]{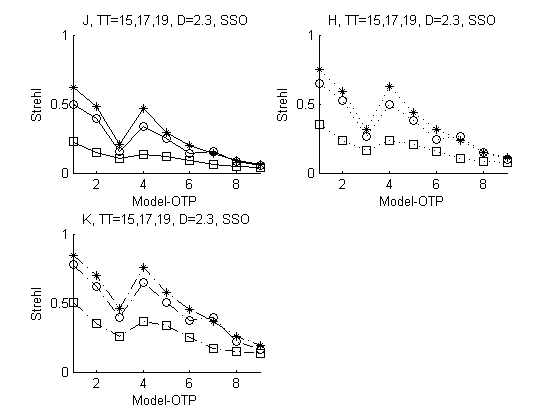}\\
  \caption[SCAO simulation results for ANU 2.3m and SSO Model-OTP for tilt guide stars of $m_R$=15,17,19 (sky coverage).] {YAO numerical SCAO simulation Strehl results for tilt guide star having $m_R$=15 (asterisks),17 (open circles),19 (open squares) for the ANU 2.3m telescope with the SSO Model-OTP (1-9). This simulation models sky coverage performance for wavelength bands J, H and K. }\label{fig:scao_aosim_tilt_sso23m}
\end{figure}

\begin{figure}[htbp]
\centering
  \includegraphics[width=1.0\columnwidth, bb=0 0  420 560]{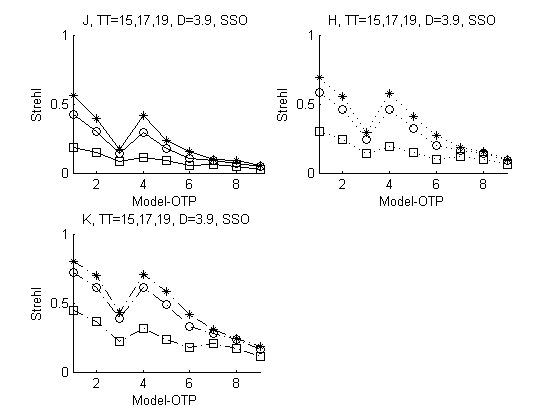}\\
  \caption[SCAO simulation results for AAO 3.9m and SSO Model-OTP for tilt guide stars of $m_R$=15,17,19 (sky coverage).] {YAO numerical SCAO simulation Strehl results for tilt guide star having $m_R$=15 (asterisks),17 (open circles),19 (open squares) for the AAT 3.9~m telescope with the SSO Model-OTP (1-9). This simulation models sky coverage performance. }\label{fig:scao_aosim_tilt_sso39m}
\end{figure}

To model the sky coverage performance, the Strehl results are shown in Figure~\ref{fig:scao_aosim_tilt_sso23m} (ANU 2.3~m ) and Figure~\ref{fig:scao_aosim_tilt_sso39m} (AAT 3.9~m) for three tip-tilt NGS of increasing limiting magnitude, $m_R$. The percentage of sky coverage for these tip-tilt NGS are tabulated in Table~\ref{tab:aosim_scao_skycoverage}.   From Figure~\ref{fig:scao_aosim_tilt_sso23m} and Figure~\ref{fig:scao_aosim_tilt_sso39m} it can be seen that both telescopes have similar trends in Strehl performance for model atmospheres 1-9. However, the AAT 3.9~m shows a marginally poorer performance in Strehl.  A noticeable drop in Strehl occurs for a tilt guide star magnitude $m_R=19$ (73\% sky coverage at 30 degrees galactic latitude) for all model atmospheres. This is unacceptable performance, particularly for the shortest wavelength J-band. It is also noted that the H-band and K-band have poor Strehls for model atmospheres 7-9  that represent a bad ground-layer resulting in the worst overall seeing conditions. Noticeable improvements the Strehl occur for H-band and K-band for model atmospheres 1 and 4, both having good free-atmosphere conditions with a good/typical ground layer.

\begin{figure}[htbp]
\centering
  \includegraphics[width=1.0\columnwidth, bb=0 0 420 560]{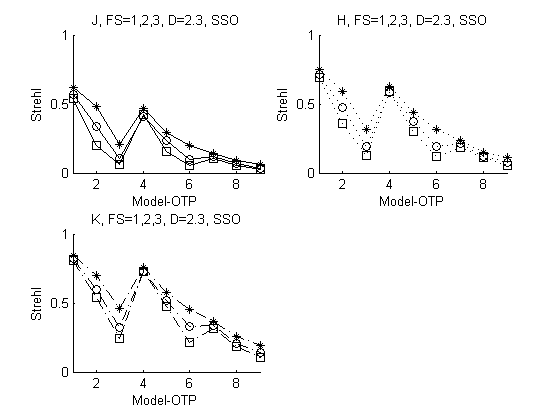}\\
  \caption[SCAO simulation results for SSO 2.~3m and SSO Model-OTP for several field stars (anisoplanatism).] {YAO numerical SCAO simulation Strehl results for field stars having $m_R$=15 and angular distance from LGS (x,y) in arcsecs of (0,0) (asterisks), (0,10) (open circles), (20,20) (open squares) for the ANU 2.3~m telescope with the SSO Model-OTP (1-9). This simulation models anisoplanatism (correction FOV) performance for wavelength bands J, H and K. }\label{fig:scao_aosim_fs_sso23m}
\end{figure}

\begin{figure}[htbp]
\centering
  \includegraphics[width=1.0\columnwidth, bb=0 0 420 560]{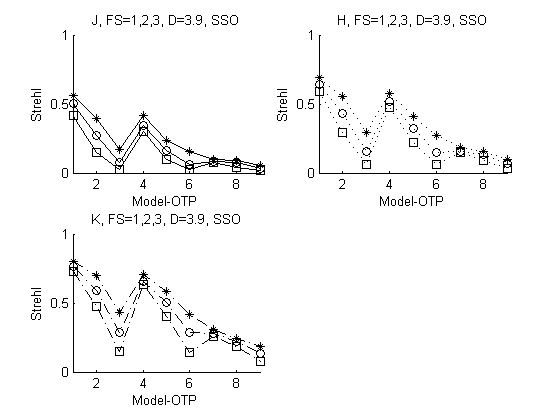}\\
  \caption[SCAO simulation results for ANU 2.3m and SSO Model-OTP for several field stars (anisoplanatism).] {YAO numerical SCAO simulation Strehl results for field stars having $m_R$=15 and angular distance from LGS (x,y) in arcsecs of (0,0) (asterisks), (0,10) (open circles), (20,20) (open squares) for the AAT 3.9~m telescope with the SSO Model-OTP (1-9). This simulation models anisoplanatism (correction FOV) performance. }\label{fig:scao_aosim_fs_sso39m}
\end{figure}

To model the anisoplanatism (correction field of view) performance the Strehl results are shown in Figure~\ref{fig:scao_aosim_fs_sso23m} (ANU 2.3~m ) and Figure~\ref{fig:scao_aosim_fs_sso39m} (AAT 3.9~m)  for three field stars of increasing angular distance from LGS (science target). From Figure~\ref{fig:scao_aosim_fs_sso23m} and Figure~\ref{fig:scao_aosim_fs_sso39m} we note similar trends in performance of both telescopes. The Strehl decreases gradually, but almost identical, for increasing angular distance of the field stars. The Strehl increases for the longer wavelengths, with highest values for the K-band.  Noticeable improvements the Strehl occur for model atmospheres 1 and 4 (good free-atmosphere conditions). The results indicate that relatively large correction fields of view are possible, $\sim 30$''. 

\subsection{MCAO Results}

\begin{figure}[htbp]
\centering
  \includegraphics[width=1.0\columnwidth]{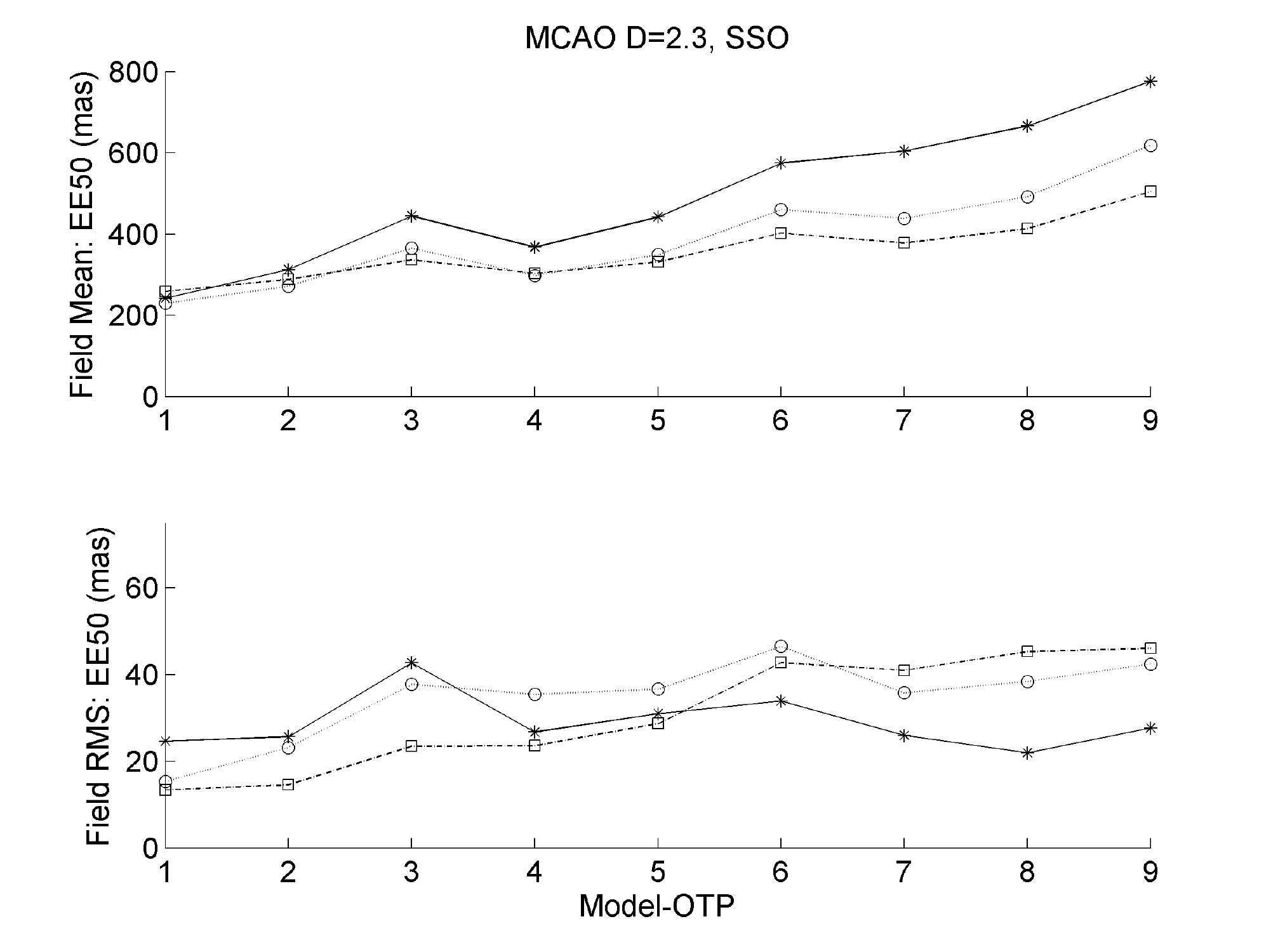}\\
  \caption[MCAO simulation results summary for ANU 2.3m and SSO Model-OTP.] {MCAO simulation results summary (using YAO) for EE50d parameter (units of milli-arcseconds) for field mean and field RMS at J-Band (asterisks), H-band (open circles), K-band (open squares) wavelengths for the ANU 2.3~m telescope with the SSO Model-OTP (1-9). }\label{fig:MCAO_EE50_Stats_fig_1}
\end{figure}

\begin{figure}[htbp]
\centering
  \includegraphics[width=1.0\columnwidth]{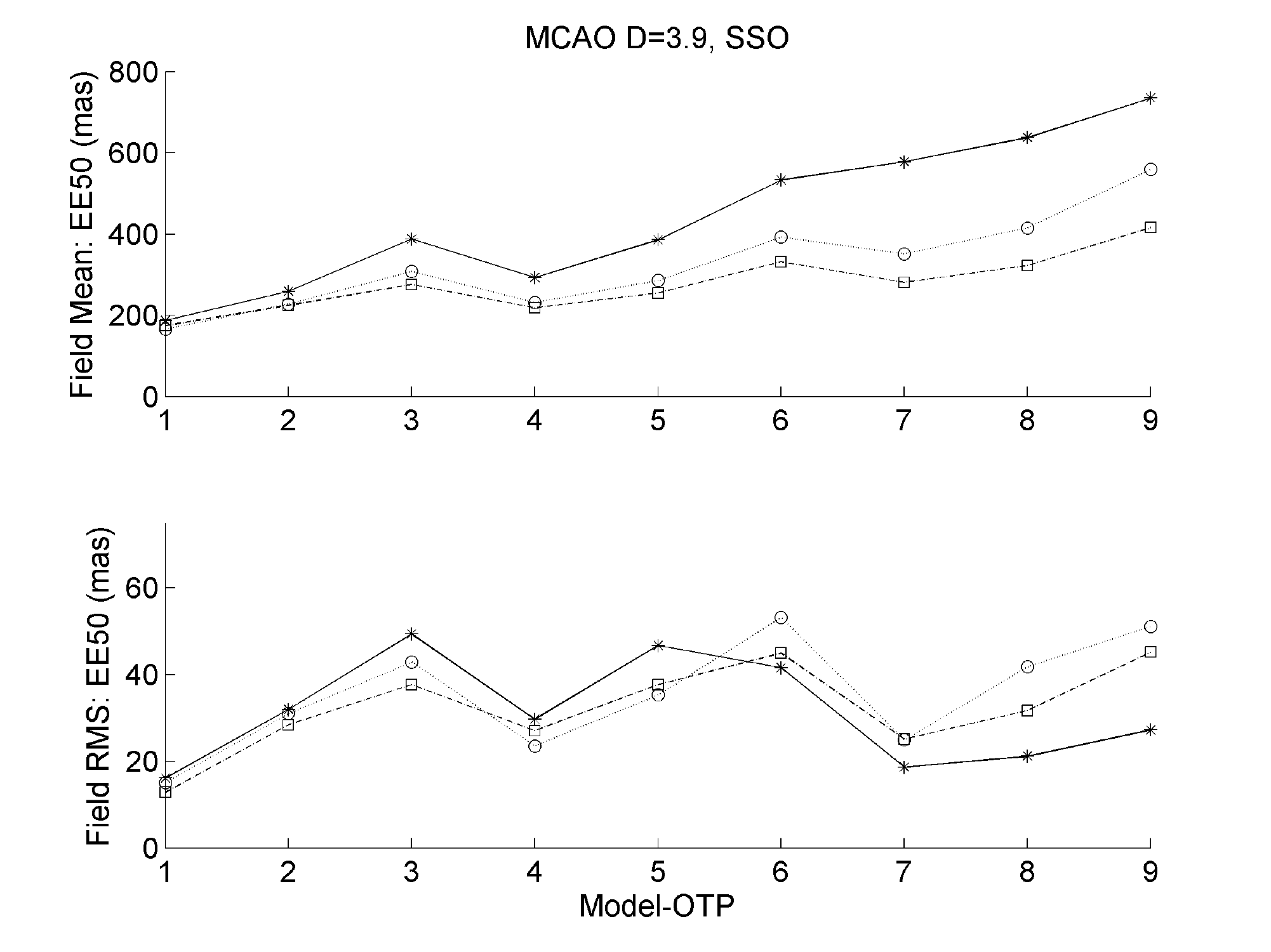}\\
  \caption[MCAO simulation results summary for AAO 3.9m and SSO Model-OTP.] {MCAO simulation results summary (using YAO) for EE50d parameter (units of milli-arcseconds) for field mean and field RMS at J-Band (asterisks), H-band (open circles), K-band (open squares) wavelengths for the AAT 3.9~m telescope with the SSO Model-OTP (1-9). }\label{fig:MCAO_EE50_Stats_fig_2}
\end{figure}

\begin{figure}
\resizebox{1.0\textwidth}{!}{\rotatebox{90}{\includegraphics[]{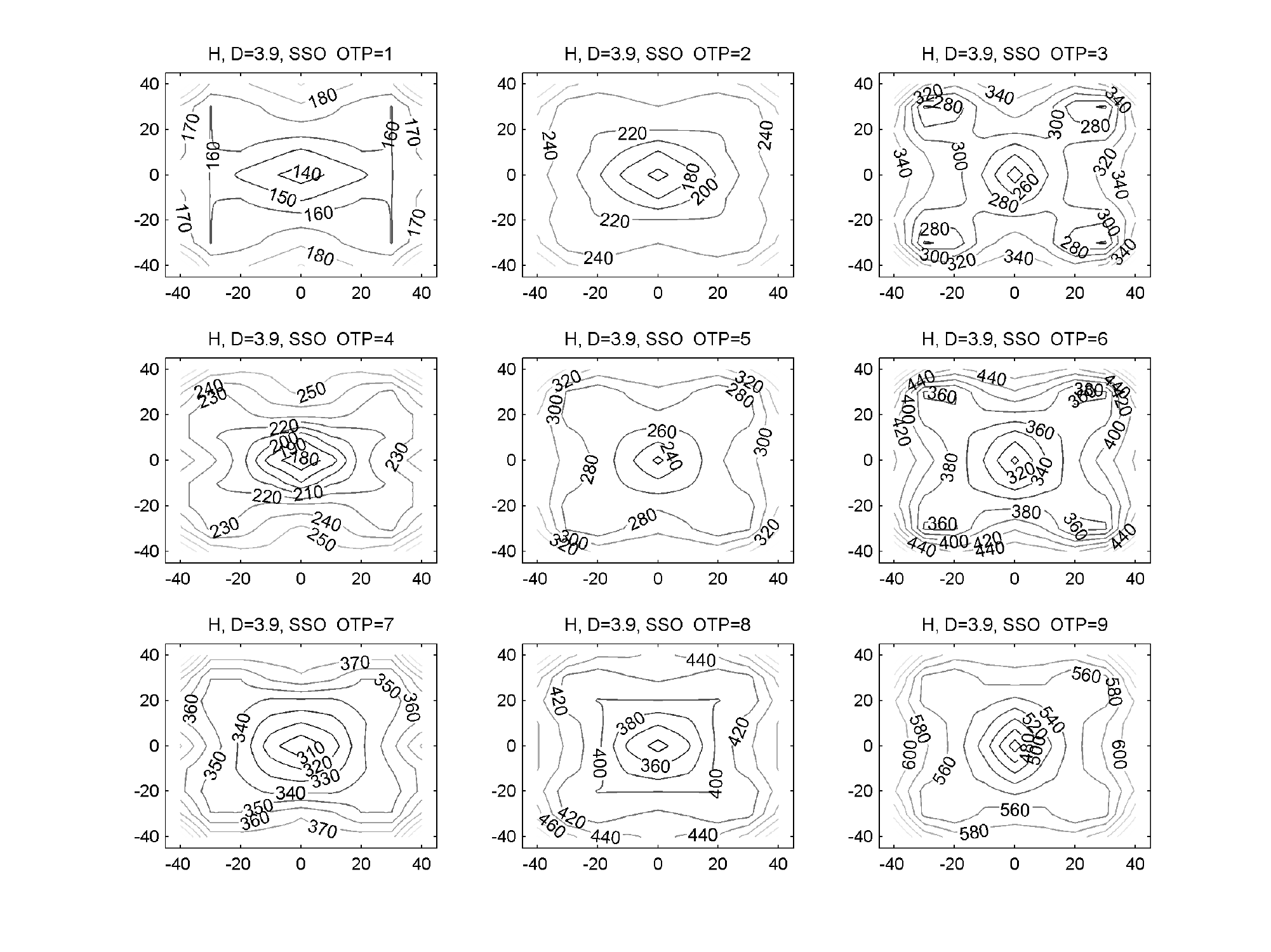}}}
  \caption[MCAO simulation results for AAO 3.9m and SSO Model-OTP shown as EE50 contour plots for H-band.] {MCAO simulation results (using YAO) for EE50 parameter (units of milli-arcseconds) at H-Band wavelengths for the AAO 3.9m telescope with the SSO Model-OTP (1-9). The $x$ and $y$ axis of the contour plots denote angular field of view (units of arcseconds). }\label{fig:MCAO_contour_ModelOTP_2_WL_2}
\end{figure}

The input MCAO configuration parameters as used in the YAO simulation code for the ANU 2.3~m and AAT 3.9~m are tabulated in Table~\ref{tab:aosim_mcao_parameters} with a schematic of the geometry shown in Figure~\ref{fig:sim_config_geometry}. The EE50d parameter (units of milli-arcseconds) is used as the figure of merit to assess the performance of the MCAO simulations. A representative contour plot of the EE50d over the designed angular field of view for the  AAT 3.9~m telescope with the SSO Model-OTP (1-9) for H-band  is shown in Figure~\ref{fig:MCAO_contour_ModelOTP_2_WL_2}. The nominal correction field of view for MCAO simulations is 80 arcseconds. A summary of MCAO simulation results for EE50d parameter's field mean and field RMS for the ANU 2.3~m and AAT 3.9~m are shown in Figure~\ref{fig:MCAO_EE50_Stats_fig_1} and  Figure~\ref{fig:MCAO_EE50_Stats_fig_2}.

 From Figure~\ref{fig:MCAO_EE50_Stats_fig_1} and  Figure~\ref{fig:MCAO_EE50_Stats_fig_2} we note similar trends in MCAO performance of both telescopes.  The results for MCAO simulations for SSO having FOV of 80 arcsecs, show sensitivity to the ground-layer turbulence with the poorest EE50d results for `bad' ground-layer conditions (poorest seeing), particularly for the shortest wavelength, or J-band. Reasonable MCAO results for SSO are achievable for longer wavelengths of H-band and K-band, with field mean of EE50d between 200 to 400~mas and corresponding field RMS between 20 to 50~mas. Conditions of strong free-atmosphere turbulence (model atmospheres 3, 6 and 9) increases both the mean and RMS of EE50d, but the MCAO sensitivity to the free-atmosphere is somewhat less than that for the SCAO performances.  The EE50d contour plot  shown in Figure~\ref{fig:MCAO_contour_ModelOTP_2_WL_2} for bad free-atmosphere conditions show the best corrections around the placement of the 5 LGS (hence large field RMS of EE50d).

\subsection{GLAO Results}

\begin{figure}[htbp]
\centering
  \includegraphics[width=1.0\columnwidth]{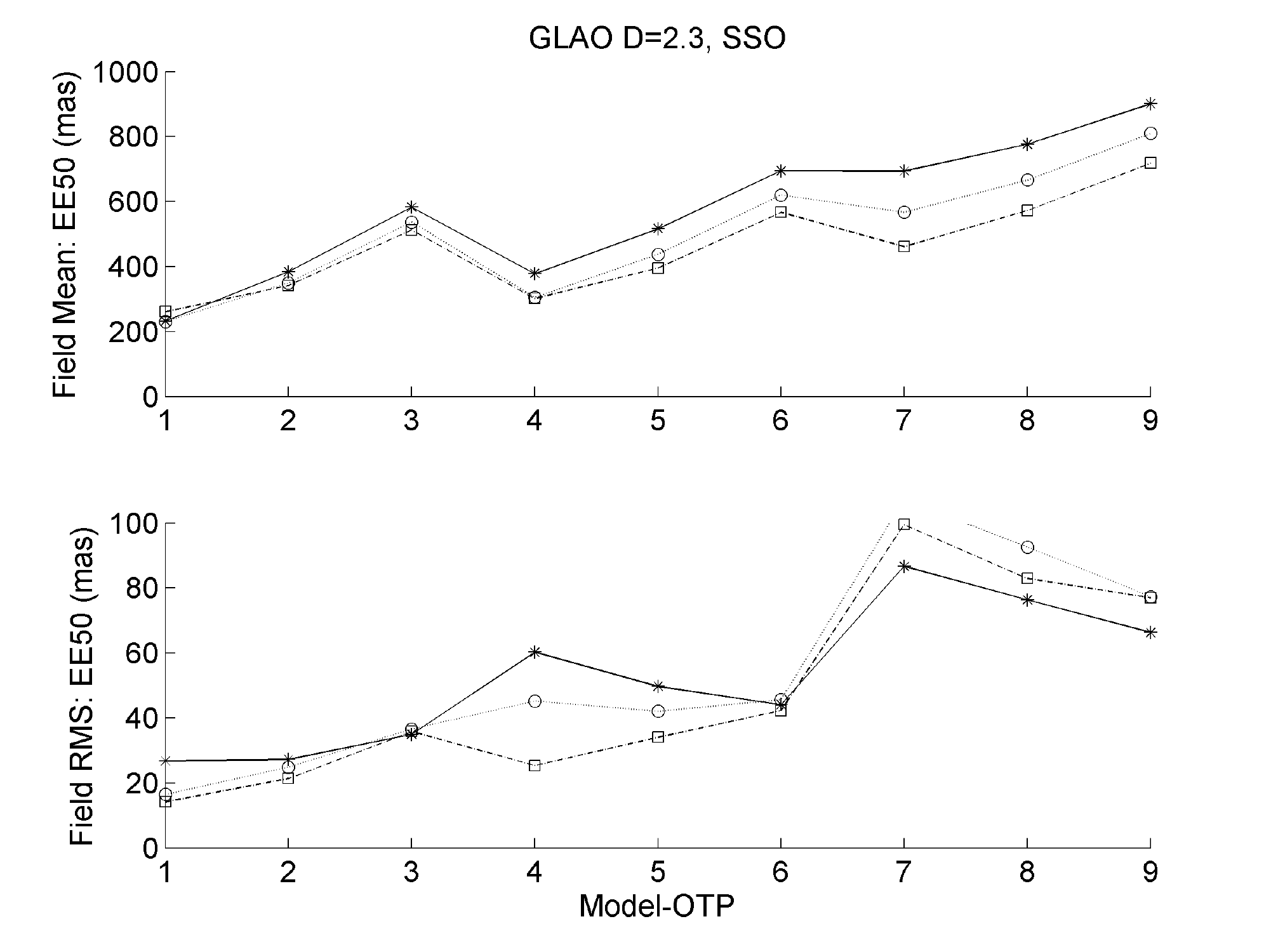}\\
  \caption[GLAO simulation results summary for ANU 2.3m and SSO Model-OTP.] {GLAO simulation results summary (using YAO) for EE50d parameter (units of milli-arcseconds) for field mean and field RMS at J-Band (asterisks), H-band (open circles), K-band (open squares) wavelengths for the ANU 2.3~m telescope with the SSO Model-OTP (1-9). }\label{fig:glao_EE50_Stats_fig_1}
\end{figure}

\begin{figure}[htbp]
\centering
  \includegraphics[width=1.0\columnwidth]{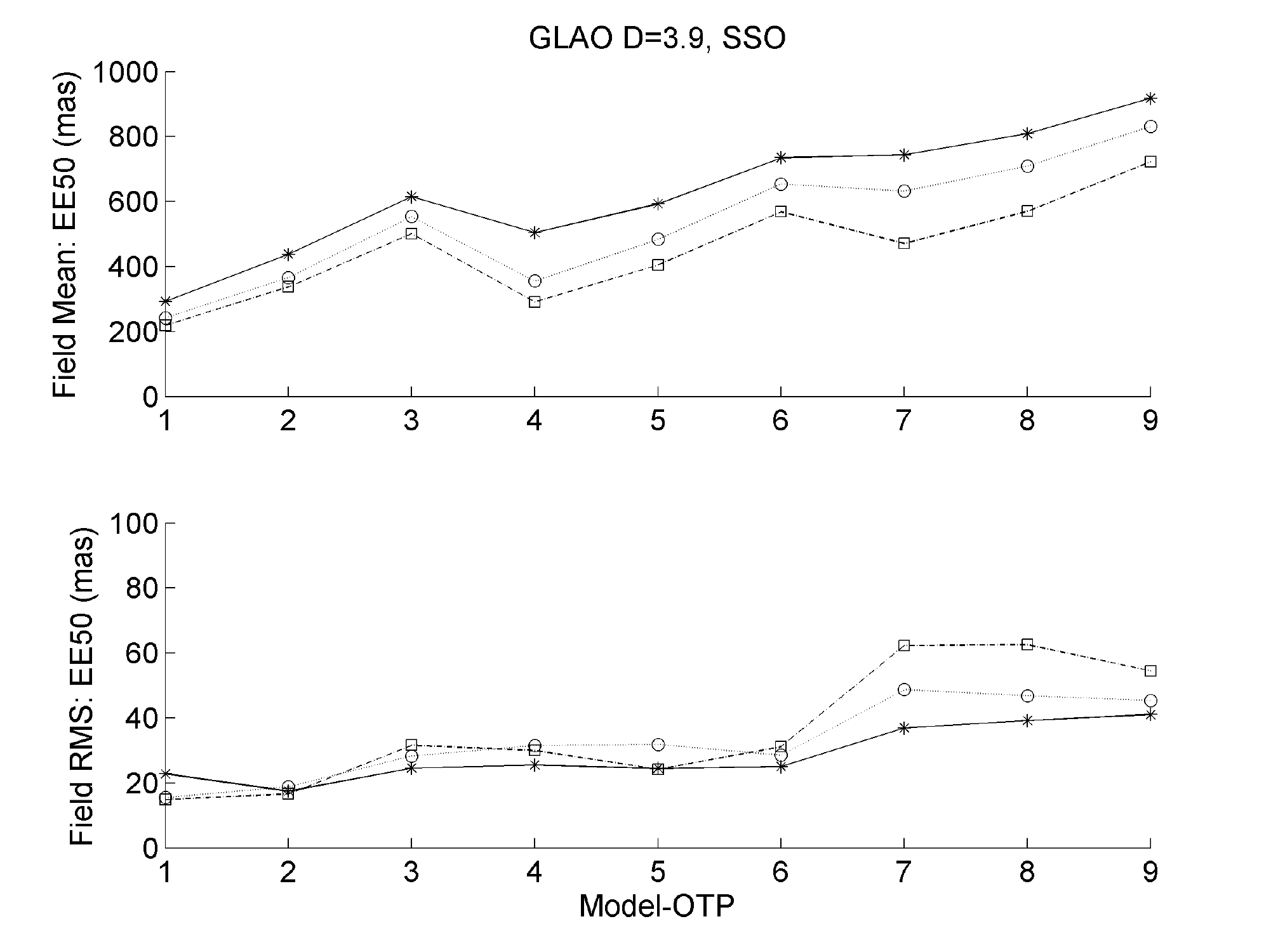}\\
  \caption[GLAO simulation results summary for AAT 3.9~m and SSO Model-OTP.] {GLAO simulation results summary (using YAO) for EE50 parameter (units of milli-arcseconds) for field mean and field RMS at J-Band (asterisks), H-band (open circles), K-band (open squares) wavelengths for the AAT 3.9~m telescope with the SSO Model-OTP (1-9). }\label{fig:glao_EE50_Stats_fig_2}
\end{figure}

\begin{figure}
\resizebox{1.0\textwidth}{!}{\rotatebox{90}{\includegraphics[]{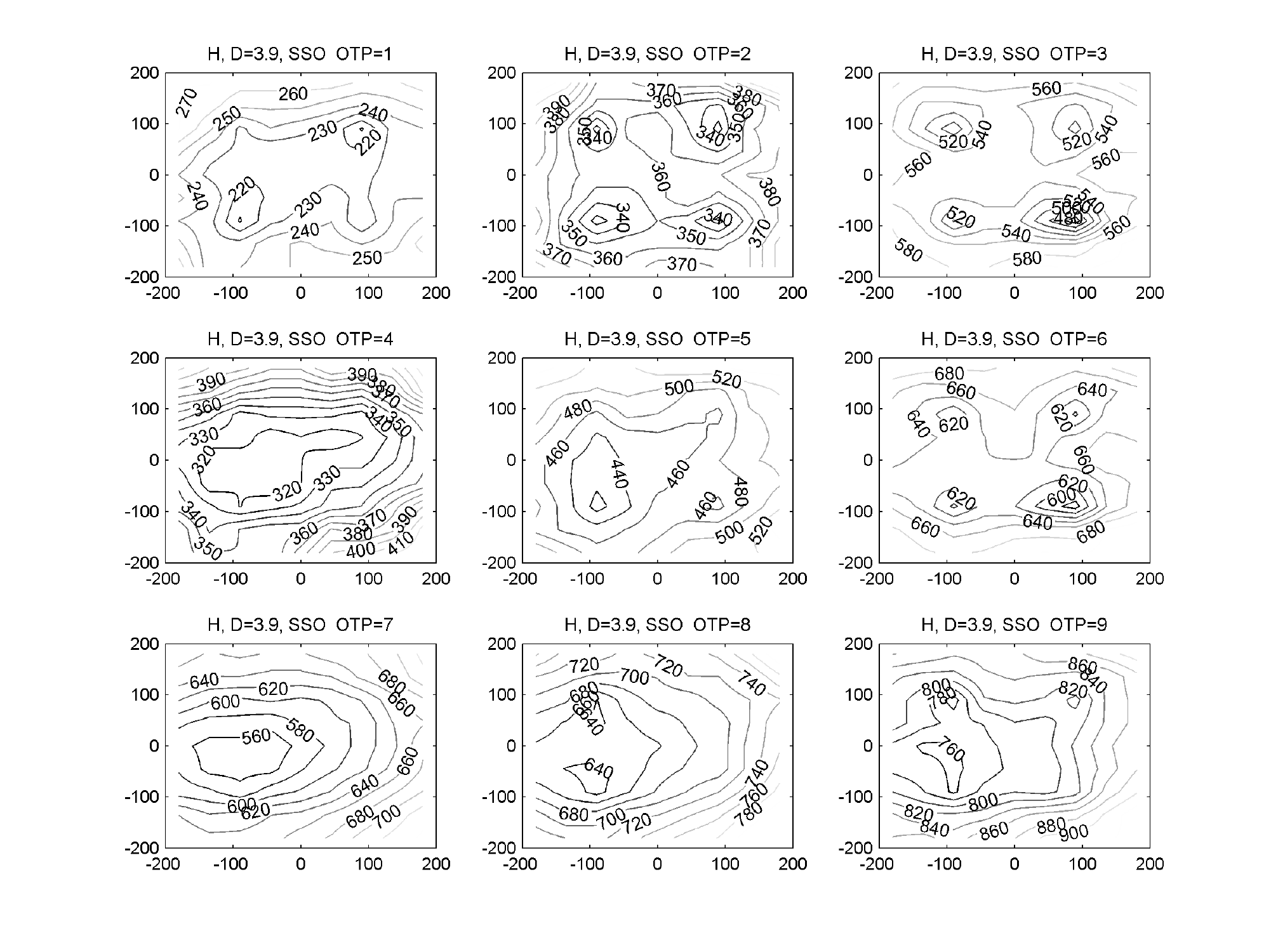}}}
  \caption[GLAO simulation results for AAO 3.9m and SSO Model-OTP shown as EE50 contour plots for H-band.] {GLAO simulation results (using YAO) for EE50 parameter (units of milli-arcseconds) at H-Band wavelengths for the AAO 3.9m telescope with the SSO Model-OTP (1-9). The $x$ and $y$ axis of the contour plots denote angular field of view (units of arcseconds). }\label{fig:glao_contour_ModelOTP_2_WL_2}
\end{figure}

The input GLAO configuration parameters as used in the YAO simulation code for the ANU 2.3~m and the AAT 3.9~m are tabulated in Table~\ref{tab:aosim_glao_parameters} with a schematic of the geometry shown in Figure~\ref{fig:sim_config_geometry}. The EE50d parameter (units of milli-arcseconds) is used as the figure merit to assess the performance of the GLAO simulations. A representative contour plots of the EE50d over the designed angular field of view for the AAT 3.9~m telescope with the SSO Model-OTP (1-9) have been simulated for wavelength  H-band (Figure~\ref{fig:glao_contour_ModelOTP_2_WL_2}). The nominal correction field of view is 180 arcseconds. A summary of GLAO simulation results for EE50d parameter's field mean and field RMS for ANU 2.3~m and AAT 3.9~m are shown in Figure~\ref{fig:glao_EE50_Stats_fig_1} and Figure~\ref{fig:glao_EE50_Stats_fig_2}.

From Figure~\ref{fig:glao_EE50_Stats_fig_1} and  Figure~\ref{fig:glao_EE50_Stats_fig_2} we note similar trends in GLAO performance of both telescopes, except that the  AAT 3.9~m has a lower, more uniform field RMS for EE50d. The results of GLAO also shows a performance trend that is similar to MCAO. Performance becomes poorer at shorter wavelengths and bad free-atmosphere conditions, with  field mean of EE50d between 200 to 800~mas and corresponding field RMS between 20 to 100~mas for all model atmospheres.

The relatively good results from the GLAO simulations may need caution due to the coarse sampling of the ground layer (37.5m, 250m and 1000m) may be insufficient to accurately model the behaviour of a 3' FoV system. The coarse sampling can lead to obtaining a GLAO result that is  too optimistic. A test to split the first few layers into many would help to verify that sampling is sufficient. This could not be fully explored due to simulation time constraints (more layers increase simulations times) and that the model is kept constant for consistent comparison of adaptive optic correction modes.

\section{Conclusions}

This paper has reported on our adaptive optic simulations for ANU 2.3~m and AAT 3.9~m telescopes based on the YAO simulation code.  A summary of results has been presented for SCAO, MCAO and GLAO based on a model-OTP derived from measurements at SSO spanning years 2005 to 2006. The results indicate that adaptive optics performance is best for the longer  wavelengths (K-band) and in the best seeing conditions (sub 1-arcsecond). 

The results for SCAO simulations for SSO show sensitivity in performance to the turbulence strength in the free-atmosphere. For a `good' free-atmosphere the Strehl is excellent while for a `bad' free-atmosphere the Strehl is dismal. A decreasing trend in performance is also observed for an increasing ground-layer strength (poorer seeing). Results suggest anisoplanatism (distance of tip-tilt NGS from science target) has a minimal impact for FOV.   Performance improves for longer wavelengths with the best performance in the K-band. Note that the SCAO results for SSO use a powerful 10~W sodium LGS (comparable to Altair LGS for Gemini-N). The reason being that the sub-aperture sizes are smaller at SSO due to the relatively poor seeing and hence more photons are needed for wavefront sensing. We believe it to be technically feasible for SSO to implement SCAO but it would be an expensive project due to the 10~W sodium LGS. The operation of SCAO for SSO (for Strehls between 0.3 to 0.8) would be limited to larger wavelengths (e.g., K-band) with sky coverage of $\sim$35\% (galactic latitude of 30 degrees) for $\sim$50\% of nights having suitable conditions.

The results for MCAO simulations for SSO having FOV of 80 arcsecs, show sensitivity to the ground-layer turbulence with the poorest EE50d results for `bad' ground-layer conditions (poorest seeing), particularly for the shortest wavelength, or J-band. Conditions of strong free-atmosphere turbulence increases the field RMS of EE50d and hence would be unsuitable for some astronomical science cases. The cost and complexity of MCAO is significantly higher than SCAO due to the requirement of multiple LGS and NGS with associated WFS and DM. Therefore, we do not recommend MCAO as a viable option for SSO.

The results for GLAO simulations for SSO having FOV of 180 arcsec, show a similar trend and performance to that of the MCAO results. However, the GLAO field RMS for EE50d is more uniform and minimally affected by the free-atmosphere turbulence for the AAT 3.9~m. The performance is reasonably good over the wavelength bands of J, H and K. The field mean of EE50d is between 200~mas to 800~mas, which is a noticeable improvement compared to the nominal astronomical seeing (870 to 1700~mas).  GLAO has the advantage of performing active optics (static and gravity / temperature variations) and also correcting for dome seeing. GLAO is also well suited to use Rayleigh LGS that are cheaper with 'industry strength' lasers commercially available.    The implementation of GLAO for SSO is therefore technically feasible given the technical requirements of multiple NGSs (or Rayleigh LGS) over a large field radius. Therefore, we recommend GLAO as a promising option for SSO.


%
%
%

\section*{Acknowledgments} 

This work makes use of observational data and analysis provided by the author's PhD thesis research conducted at the  Research School of Astronomy and Astrophysics (RSAA) of the  Australian National University (ANU). The authors are grateful to Jon Lawrence of the Australian Astronomical Observatory (AAO) for his suggestions for the original version of the manuscript and to the referees for their comments.


\bibliographystyle{astron} 
\bibliography{references}


\onecolumn

\clearpage


\end{document}